\documentstyle[eqsecnum,aps,epsf]{revtex} 
\newcommand{\postscript}[2] {\setlength{\epsfxsize}{#2\hsize}
\centerline{\epsfbox{#1}}}

\begin{document}
\twocolumn[\hsize\textwidth\columnwidth\hsize\csname@twocolumnfalse\endcsname

\title{Positronium-Hydrogen-Atom Scattering in a Five-State Model} 

\author{Sadhan K. Adhikari and P. K. Biswas}
\address{Instituto de F\'{\i}sica Te\'orica, 
Universidade Estadual Paulista
01.405-900 S\~ao Paulo, S\~ao Paulo, Brazil\\}

\date{\today}
\maketitle

\begin{abstract}

The scattering of ortho-positronium (Ps) by hydrogen atoms has been
investigated in a five-state coupled-channel model allowing for
Ps(1s)H(2s,2p) and Ps(2s,2p)H(1s) excitations using a recently proposed
electron-exchange model potential.  The higher ($n\ge 3$) excitations and
ionization of the Ps atom are calculated using the first Born
approximation.  Calculations are reported of scattering lengths, phase
shifts, elastic, Ps- and H-excitation, and total cross sections. 
Remarkable correlations are observed between the S-wave Ps-H binding
energy and the singlet scattering length, effective range,
  and resonance energy obtained in various model calculations.  These
correlations suggest that if a Ps-H dynamical model yields the correct
result for one of these four observables, it is expected to lead to the
correct result for the other three.  The present model, which is
constructed so as
 to reproduce the Ps-H resonance at 4.01 eV, automatically yields a Ps-H
bound state at $-1.05$ eV which compares well with the accurate value of
$-1.067$ eV. The model leads to a singlet scattering length of 3.72$a_0$
and effective range of 1.67$a_0$, whereas the correlations suggest the
precise values of $3.50a_0$ and 1.65$a_0$ for these observables,
respectively.

{\bf PACS Number(s):  34.10.+x, 36.10.Dr}

\end{abstract}

\vskip1.5pc]

\newpage

\section{Introduction}

Recently, there have been several experimental and theoretical investigations
of ortho positronium (Ps)  atom scattering from different neutral atomic and
molecular targets. Experiments have primarily measured  total Ps-atom
scattering cross sections from various targets \cite{2,3,4}.  In addition to
total cross sections, the theoretical studies have also predicted partial cross
sections and phase shifts for Ps-H \cite{5,5a,6,7,8,ba1}, Ps-He
\cite{8,9,9a,ba2},
 and Ps-H$_2$ \cite{10,ba3} systems.  Ps scattering by neutral
targets is of special interest, as the direct amplitudes for elastic and
even-parity state
transitions are
zero \cite{ba4} 
due to the internal charge and mass symmetry of Ps. Hence the 
electron-exchange interaction is the dominating factor at low energies in any
Ps-impact scattering with neutral targets apart from the effect of 
polarization and van der Waals forces \cite{9,gh}.  Ps scattering makes it
possible to study
  the effect of exchange  in an environment
characteristically different from that of the electron-atom systems
\cite{ba4}.  Among
all Ps-atom systems, the positronium-hydrogen (Ps-H) system is the simplest
and is of fundamental interest.
The Ps-H scattering has most of the 
complications of a many-body problem, but few-body techniques can be employed
for its solution.

A general feature of the measured  total cross section in Ref.
\cite{2} for Ps scattered by  He, Ar, and H$_2$, is a peak near 20 $-$
25 
eV and a decreasing trend
below this energy. Recent measurements near 1 eV 
\cite{3} are consistent  with this trend. 
However, because of the  large error bars 
on the measurement in Ref. \cite{2} at the lowest energy (10 eV)
and due to inadequate data in this energy region, it is 
not clear from experiment whether the total cross section has a minimum
near the Ps excitation threshold or not. The recent three-Ps-state 
studies of Refs. \cite{ba1,ba2,ba3}  suggest the existence of a minimum near
the Ps(2s)  threshold.  This feature of the cross section is able to
reconcile the two different experimental findings and is also noticed in the
unpublished theoretical work of Peach \cite{gp}.  The 
 R-matrix calculation
\cite{7,8} for H and He, in which 22 coupled pseudo states are included,
does not show this trend;  
 whereas the static-exchange
(one-state) cross sections \cite{9} for He are too large to match the
measurement near the Ps(2s) threshold \cite{ba2}.  In this respect, the
model-potential studies of Refs. \cite{ba1,ba2,ba3} are unique in reproducing
the experimental trend of Ps-impact scattering by  H, He, and H$_2$.
Unphysically large low-energy 
cross sections of previous cross sections are expected to
be a consequence of the nonorthogonality  arising from  
antisymmetrization 
 and the very reactive nature of Ps scattering
\cite{ba1,ba2,ba3}.

In this paper we present a theoretical study of ortho-Ps-H scattering
 employing a five-state model allowing for excitation of
both Ps and H atoms using the model exchange potential mentioned above.
  The following states are included in the calculation:
Ps(1s)H(1s), Ps(2s)H(1s), Ps(2p)H(1s), Ps(1s)H(2s), and Ps(1s)H(2p)
 and such a model should be
considered adequate  at low energies.  The cross sections for higher discrete
and continuum excitations of the Ps atom are calculated in the framework of
Born approximation. These Born cross sections are added to the
above five-state  cross sections to predict the total  cross
section.

Although,
the parametrization of the model exchange potential of Ref. \cite{ba2},
which is obtained using a physical argument,  is satisfactory, it   
 is not unique and leaves an
option for the parameters to be varied to tune to some precise
data at low energies. In the absence of experimental Ps-H cross
sections we tune this parameter to reproduce  the energy of 
the singlet S-wave   resonance.  
Ho has provided the most  precise estimate of S-wave resonance energy,
which is 
 4.01 eV (width 0.075 eV) \cite{ho2}.  
Frolov and Smith  have
made 
the most accurate
estimate of the S-wave  bound state, which is 1.067 eV \cite{ho1}.
We varied
the parameter of our model to fit the Ps-H resonance energy at 4.01 eV and
found that the same model  without any further  adjustment  also produced a
Ps-H bound state at  $-1.05$ eV. No previous scattering model has been able
to produce these two features of the Ps-H system simultaneously and 
so precisely. Similar to those obtained in the three-nucleon system
\cite{ad,ph}, we find
remarkable correlation between the S-wave Ps-H binding
energy
and 
the singlet  scattering
length, effective range,
  and  resonance energy obtained in various model calculations. 
These correlations 
 suggest that if
a model yields the correct result for  one of these observables it should
also yield the correct result for the other three. 
The present model leads to reasonably accurate energies 
 for the  Ps-H bound state and
resonance and because of the above  correlation
the singlet scattering length and effective range are 
also expected to
be fairly accurate.

We describe the calculational scheme, model exchange potential  
and numerical results in Sec. II and a summary of our findings in Sec.
III.

\section{Model Positronium-Hydrogen Calculation}

\subsection{Calculational Scheme}

The total antisymmetrized wave function for the Ps-H system allowing 
excitation of both Ps and H is given by
\begin{eqnarray}\label{a}
\Psi^\pm({\bf r}_1,{\bf r}_2, {\bf x}) & \equiv& \frac{1}{\sqrt 2}\sum_{\mu,\nu}
[\phi_\mu ({\bf r}_ 2)  \chi_\nu ({\bf t}_ 1)  
F_{\mu \nu} ({\bf s}_ 1)\nonumber \\
&\pm & 
\phi_\mu ({\bf r}_ 1) \chi_\nu ({\bf t}_ 2) F_{\mu \nu} ({\bf s}_ 2)], 
\end{eqnarray}
where ${\bf s}_j = ({\bf x +r}_j)/2$   and ${\bf t}_j = ({\bf x -r}_j)$, 
$j=1,2$,
with ${\bf x}$ the positron coordinate and ${\bf r}_j$ are the coordinates of
the two electrons, $\phi_\mu$ ($\chi_\nu$) is the bound-state wave function 
of H (Ps) in quantum state $\mu$ $(\nu)$, and $F_{\mu \nu}$ is the continuum 
orbital of Ps with respect to H. The Schr\"odinger equation for this 
 wave function when projected on the final H and Ps states
  $\phi_{\mu ' }$ and 
$\chi_{\nu '}$, respectively, leads to the following  
Lippmann-Schwinger scattering integral
equation in momentum space
\begin{eqnarray}
f^\pm_{\mu '\nu ',\mu\nu} ( {\bf k',k})&=&
{\cal B}^\pm _{\mu '\nu ',\mu\nu}({\bf k ',k})
\nonumber
\\
&-&\sum_{\mu ",\nu "}
\int \frac{   {\makebox{d}{\bf k"}}    }   {2\pi^2}  
\frac {   {\cal B}^ \pm 
 _ {\mu '\nu ',\mu" \nu"}  
({\bf k ',k"}) 
f^ \pm                   _{\mu" \nu" ,\mu\nu}      ({\bf k",k}) }
{E-{\cal E}_{\mu "}-\epsilon_{\nu "}-k"^2/4+ \makebox{i}0}.
\nonumber \\ \label{4}
\end{eqnarray}
where the
singlet and triplet ``Born" amplitudes, $B^\pm$,   are given by  $
 {\cal B}^\pm_{\mu '\nu ',\mu\nu}({\bf k_f,k_i}) = 
 g^D_{\mu '\nu ',\mu\nu}({\bf
k_f,k_i})\pm g^ E_{\mu '\nu ',\mu\nu}({\bf k _f,k_i}),$ 
   where $g^D$  and $g^E$ represent the direct and 
exchange Born  amplitudes and the $f^ \pm$ are 
the singlet and triplet  
scattering amplitudes, respectively. The energies of the intermediate Ps
and H states
are
$\epsilon_{\nu''} $  and ${\cal E}_{\mu '' }$ 
 and $E$ is the total
energy of the system.
  The differential cross section is defined 
by 
\begin{equation}
\left(\frac{d\sigma}{d\Omega}\right)_{\mu '\nu ',\mu\nu} = \frac
{k'}{4k }
[|f^ +_{\mu '\nu ',\mu\nu}({\bf k',k})|^2+
3|f^ -_{\mu '\nu ',\mu\nu}({\bf k',k})|^2   ],
\end{equation}
and the quenching cross section that describes conversion from ortho- to 
para-positronium is defined
by 
\begin{equation}\left(\frac{d\sigma}{d\Omega}\right)_{\mu '\nu ',\mu\nu}^
{\makebox{quen}} = \frac {k'} {16k}
|f^+_
{\mu '\nu ',\mu\nu}({\bf k',k}) -f^- _{\mu '
\nu ',\mu\nu}({\bf k',k})|^2.
\end{equation}

\subsection{Model Exchange Potential}

The derivation of the model exchange  potential has been adequately
described recently and here we quote the principal results
\cite{ba2}.  The Ps-H model exchange potential is given by \cite{xx}
\begin{eqnarray}\label{1} 
g^E_{\mu'\nu',\mu\nu} ({\bf k_f,k_i})&= &
\frac{4(-1)^{l+l'}}{D}
\int \phi^*_{\mu '}({\bf r})\exp ( \makebox{i} {\bf Q. r})\phi_\mu({\bf r})
\makebox{d}{\bf r}_2\nonumber \\ &\times&
\int \chi^*_{\nu '}({\bf  t })\exp ( \makebox{i}{\bf Q}.{\bf  t }/
2)\chi_{\nu}({\bf  t } ) \makebox{d}{\bf  t },
\end{eqnarray}
with 
\begin{equation}\label{2}
D={k_f^2/4+\alpha_\mu^2+\beta_{\nu '}^2} 
\end{equation}
where $l$  and $l'$ are  the angular momenta of the initial and final Ps
states, 
the initial and 
final Ps momenta are ${\bf k_i}$  and ${\bf k_f}$, ${\bf Q = k_i -k_f}$.
$\alpha_\mu^2/2$ and $\alpha_{\mu ' }^2/2$, and $\beta_\nu^2$
and $\beta_{\nu '}^2$ are the binding energies of the initial and final
states of  H and  Ps in atomic units (au), 
respectively.  The factor  $(-1)^{l+l'}$ provides 
the correct sign of the exchange potential given by formal
antisymmetrization.  In previous works \cite{ba1,ba2,ba3} only the $l=0$ 
component of this model potential was given.
The model exchange potential  given by Eqs. (\ref{1}) and (\ref{2})  is
not time-reversal symmetric. A time-reversal symmetric form has also been
suggested in which Eq. (\ref{2}) is replaced by 
\cite{ba2}:
\begin{equation}\label{3}
D={(k_i^2+k_f^2)/8+(\alpha_\mu^2+\alpha_{\mu '}^2)/2+(\beta_\nu^2+
\beta_{\nu ' }^2)/2}, 
\end{equation}
which leaves the elastic Born results unchanged.

\subsection{Numerical Results}

After a partial-wave projection, the singlet (+) and triplet ($-$) 
scattering equations  (\ref{4})  are solved by the
method of matrix inversion. The maximum number of partial waves included
in the
calculation is ten.  Contribution of higher partial waves to cross
sections
is included by  using the  Born terms.

\vskip -3.5cm
\postscript{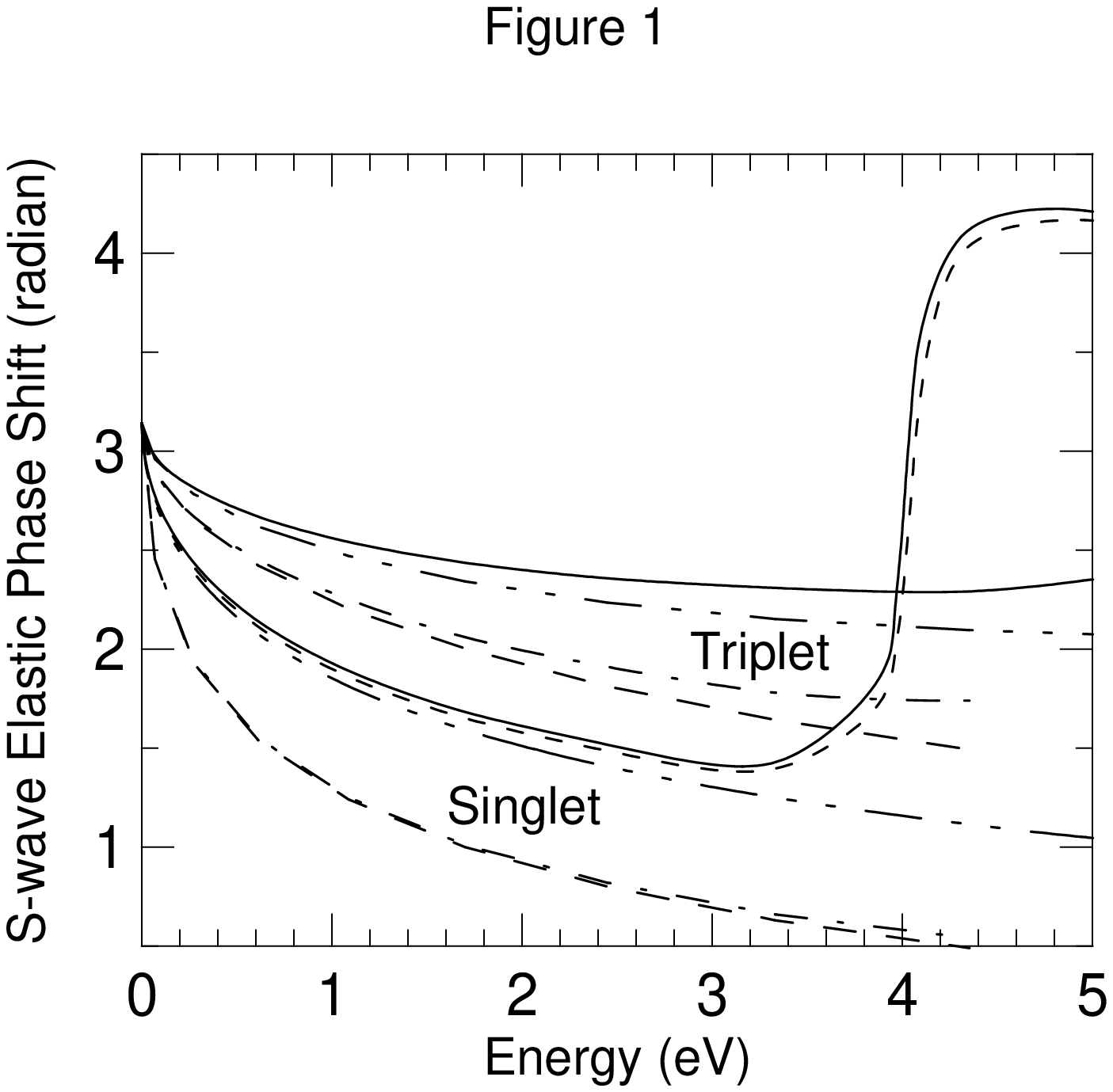}{1.0}    
\vskip -1.9cm
\noindent {\small {\bf Fig. 1} \ \
 S-wave elastic scattering phase shifts for  singlet and triplet
states at different Ps energies: present five-state 
 (full line), present two-state 
 (dotted line), 
 present static exchange (dashed-double-dotted line)
Hara and
Fraser (dashed line, Ref. \cite{5a}), Sinha et al. (dashed-dotted line, Ref.
\cite{6}).}

\vskip -2.4cm
\postscript{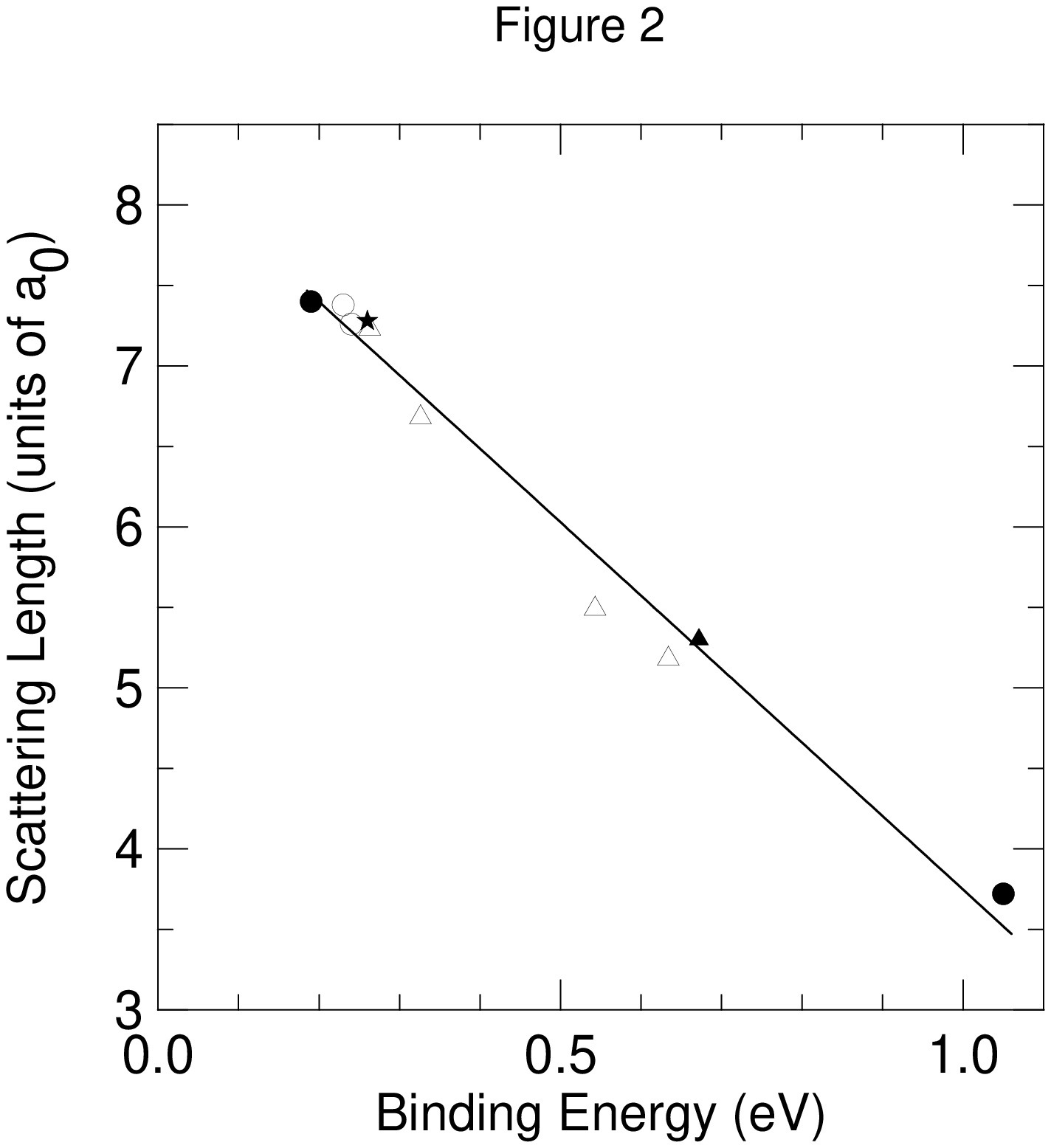}{1.0}    
\vskip -2.3cm
\noindent {\small {\bf Fig. 2} \ \
The singlet scattering length versus binding energy of different
models: open triangles (Ref. \cite{7}), open circles (calculated from phase 
shifts of Ref. \cite{6}), solid triangle (from Ref. \cite{5}), star
(as calculated in Ref. \cite{5} from  phase shifts of \cite{5a}), 
solid circle (five-state 
calculation with $C$ = 1 and 0.784 in Eq. (\ref{5})),
full line (straight line fit).}

In our latest calculations \cite{ba2} we find that the symmetric form 
provides better results and therefore
here we present results of Ps-H scattering using a five-state model and 
 Eqs. (\ref{1}) and (\ref{3}) that includes the 
 following states: Ps(1s)H(1s), Ps(2s)H(1s),
Ps(2p)H(1s), Ps(1s)H(2s), and Ps(1s)H(2p).  The truncated model
that includes the
first three of these states will be referred to as the three-Ps-state, or
simply, the three-state, model, and the three-H-state model 
 includes the first, fourth and fifth of
this set. The model  that includes the
first
$n$ states of this set will be termed the $n$-state model.  The Born terms
for the  simultaneous excitation of both H and Ps atoms are found to be 
 small and will not be
considered here in the coupled-channel scheme.  
Higher excitations and ionization of Ps  are conveniently
treated  in the Born approximation including exchange and   
higher excitations
and ionization of H  are excluded.  We calculate the elastic
Ps(1s)H(1s)  cross section and inelastic excitation cross sections to
Ps(2s)H(1s), Ps(2p)H(1s), Ps(1s)H(2s), and Ps(1s)H(2p) states. We also
calculate cross sections for the discrete excitation 
of the 3s, 3p, 3d, 4p, 4d,
4f, 5p, 5d, 5f, 6p states  and also for
ionization  of Ps in the  first Born approximation, keeping
the
target frozen to its initial ground state using  exchange given by Eqs.
(\ref{1}) and (\ref{3}).

\vskip -2.8cm
\postscript{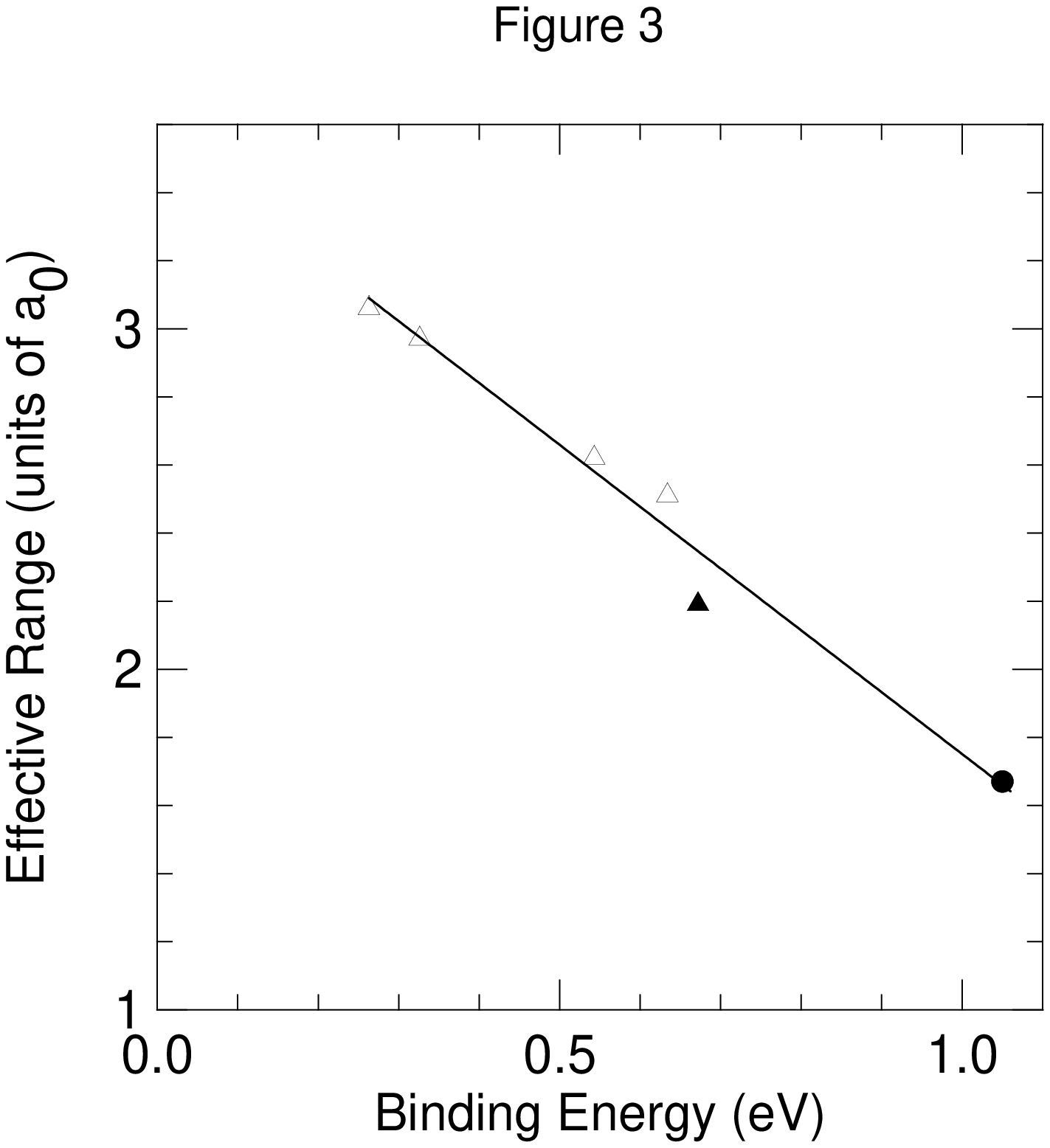}{1.0}    
\vskip -2.0cm
\noindent {\small {\bf Fig. 3} \ \
 S-wave singlet effective range  versus binding energy of different
models: open triangles (Ref. \cite{7}), 
 solid triangle (from Ref. \cite{5}), 
solid circle (five-state calculation 
with $C$ =  0.784 in Eq. (\ref{5})),
full line (straight line fit).}

\vskip -4.8cm
\postscript{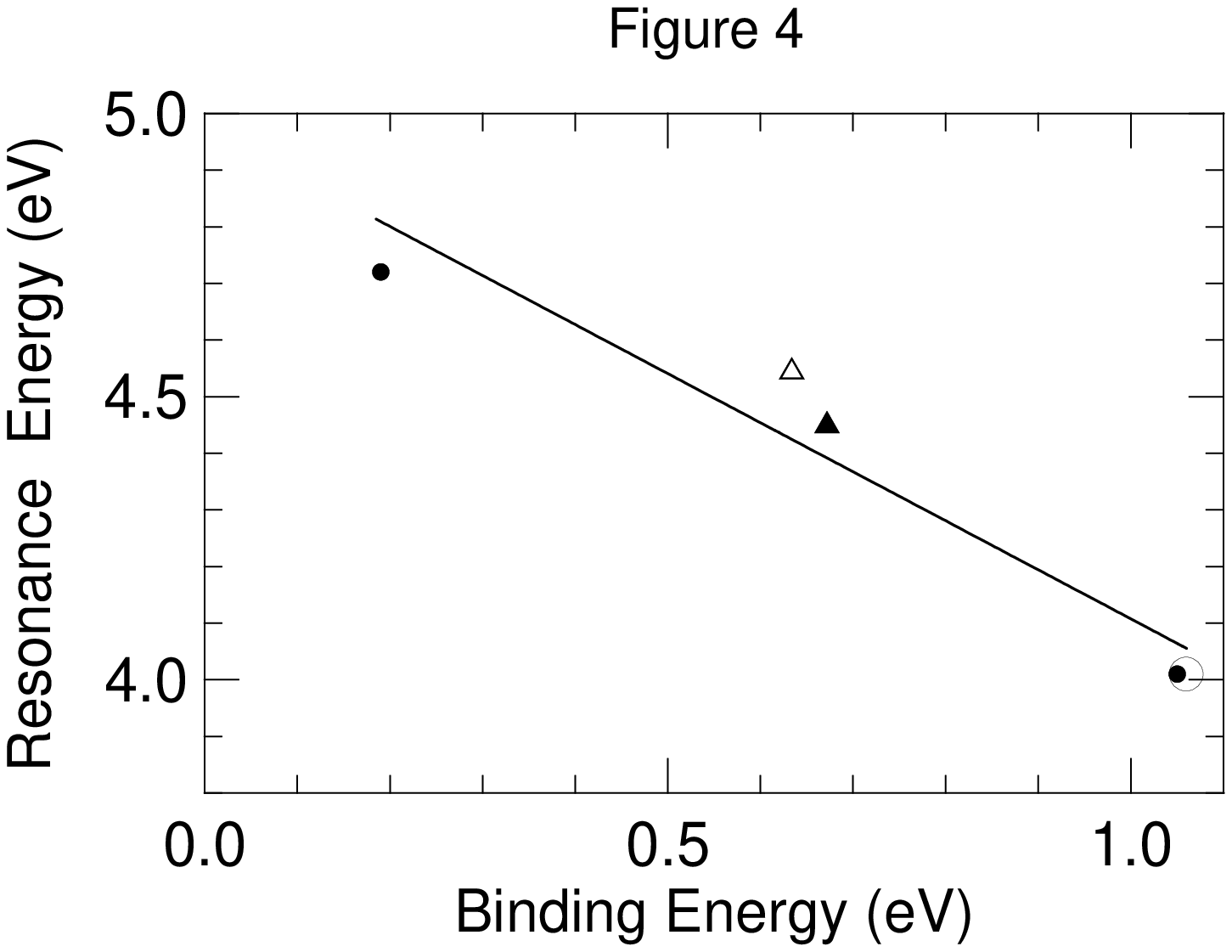}{1.0}    
\vskip -1.7cm
\noindent {\small {\bf Fig. 4} \ \
4. S-wave singlet resonance versus binding energy of different
models: open circle (Refs. \cite{ho1,ho2}),
 open triangle (Ref. \cite{7}), 
 solid triangle (from Ref. \cite{5}), 
 solid circle (five-state calculation with $C$ = 1 and 0.784 in Eq.
(\ref{5})), full line (straight line fit).  }

In previous studies we found that the exact values of the parameters
$\alpha$ and $\beta$ in Eqs. (\ref{2}) and (\ref{3}) lead to good results
for cross sections. However, these parameters in Eqs. (\ref{2}) and
(\ref{3})  correspond to some average value of momentum \cite{ba2}
and it was noted that one could conveniently allow these parameters to vary
in order to improve the fit with experiment.  As there are no experiments in
Ps-H scattering, we choose these parameters to fit the known S-wave
singlet resonance at 4.01 eV in the Ps-H system
\cite{ho2}. The resonance is found in the  two-state model with
Ps(1s)H(1s) and Ps(2s)H(1s) states. It continues to exist as more states
are included in the dynamical equation.  However, its energy reduces a
little (by up to  about 0.05 eV)  as more and more states are added. The
position of the resonance in the five-state and two-state  models with the
exact parameters $\alpha$'s and $\beta$'s in Eq. (\ref{3}) is 4.72 eV
and 4.76 eV, respectively. We find that the Ps-H resonace energy decreases
and the
binding energy
increases monotonically, 
as the values of
the parameters $\alpha$'s and $\beta$'s are reduced in Eq. (\ref{3}). For
a systematic reduction we used in place of Eq.  (\ref{3})  the following
form:  \begin{equation}\label{5}
D={(k_i^2+k_f^2)/8+C^2[(\alpha_\mu^2+\alpha_{\mu '}^2)/2+
(\beta_\nu ^2+\beta_{\nu '}^2)/2}], \end{equation} 
where $C$ is an arbitrary factor. In
order to obtain the S-wave resonance at 4.01 eV in the  five-state
model we need $C=0.784$, which is the  most accurate estimate of this
 energy  \cite{ho2}. 
Interestingly enough, with this value of $C$, the
 five-state model produces a Ps-H bound state at $-$1.05
eV, which is consistent with both the accurate 
theoretical estimate of $-$1.067 eV
\cite{ho1} and experimental result of $-1.1\pm0.2$ eV \cite{sh}.  
The
 binding energy is calculated by extrapolating the calculated $k
\cot \delta$ values at positive energies to negative energies using the
following effective-range 
expansion:  $k\cot \delta =-1/a+r_0 k^ 2/2+ B k^ 4$,
 and finding
the solution of $k\cot\delta -ik=0$ at the bound state,  where
$\delta$ is the S-wave phase shift, $a$ the scattering length, $r_0$ the
effective range, $k$ the momentum, and 
$B$  the coefficient of the $k^ 4$ term.  In all
calculations presented in this work we use the value of $C =0.784$  in
Eqs. (\ref{5})  and (\ref{1}).  The simultaneous accurate reproduction of
both the binding  and resonance energies 
assures the reliability of our
model.

The  elastic scattering S-wave phase shifts for different
partial waves for singlet and triplet states below the lowest
excitation threshold are shown in Fig. 1.
We compare the S-wave phase shifts with the static-exchange phase shift
of Hara and Fraser \cite{5a} and the three-state 
close-coupling approximation (CCA) phase shifts of Sinha
et al. \cite{9}.  We also show our static-exchange
phase shifts.    The phase shifts of Hara and
Fraser are
identical with the static exchange   results of Ref. \cite{9}.
The phase shifts of Sinha et al. and of Hara and Fraser 
 suggests that the trend of convergence of the S-wave phase shifts of
Ref. \cite{9} is in the direction of the present phase shifts.

Because of the existence of a low-energy effective-range expansion, the
binding energy of a weakly bound singlet Ps-H state should be correlated with
the S-wave singlet scattering length in different model calculations.  This
is shown in Fig. 2  where we plot the singlet scattering length
versus binding energy for  several calculations.  The straight-line
correlation
between these two observables for  various model calculations  implies
that a
model that produces the correct energy of the Ps-H bound state, should also
produce the correct scattering length and good low-energy phase shifts. This
correlation explicit in the effective-range expansion is a consequence of the
dynamics of the problem. The dominance of the short-range part of the
interaction is responsible for the appearance of correlation between 
low-energy
observables in a system \cite{ad}. 
 In the trinucleon system
in the attractive S-wave doublet channel all low-energy observables
were
found to be correlated with  binding energy in different model
calculations \cite{ad,ph}, which implies if a model yields the
correct
result for one
of the low-energy
observables it should also yield the correct result for the others. Such
correlations were used  to predict different
 low-energy trinucleon observables from results of different model
calculations. These predictions were later confirmed in other rigorous
calculations and experiments \cite{ph}.

\vskip -3.cm
\postscript{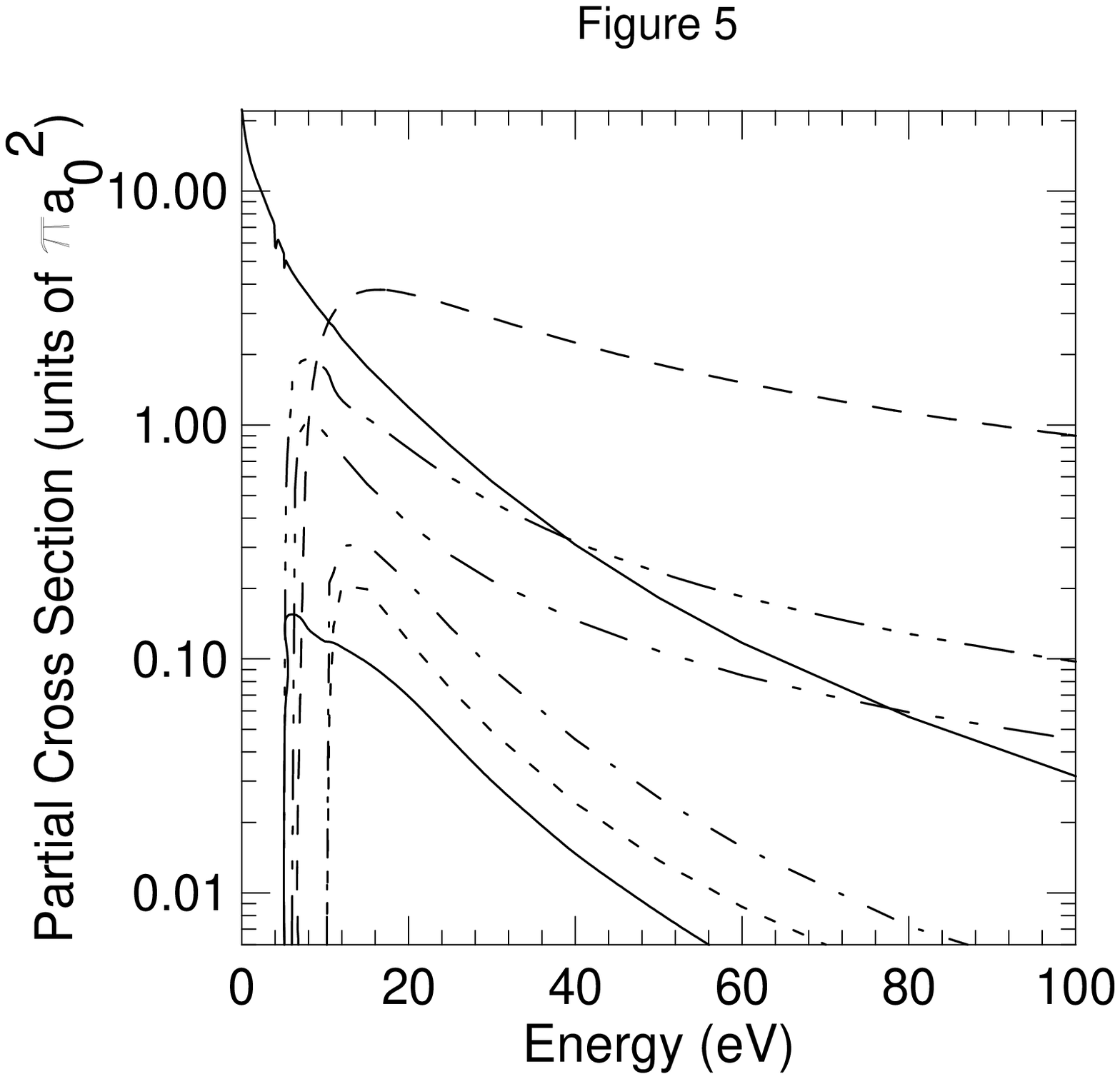}{1.0}    
\vskip -1.7cm
\noindent {\small {\bf Fig. 5} \ \
Partial  Ps-H cross sections from the five-state 
model at different Ps
energies: Ps(1s)H(1s) elastic (upper full  line),
Ps(2s)H(1s) excitation (lower full line),
Ps(1s)H(2s) excitation (dotted line), 
Ps(1s)H(2p) excitation  (dashed-dotted line),
Ps(2p)H(1s) excitation  (dashed-triple-dotted  line),
 Ps-ionization Born cross section (dashed line),
Ps-excitation ($n\ge 3$) Born cross section
(dashed-double-dotted line).}

Correlation is also possible among other
low-energy S-wave singlet Ps-H 
observables which are not  obviously related.
For example, we find a  correlation between  S-wave singlet Ps-H
binding energy and  effective range, which is shown in Fig. 3. In this figure
we plot the effective ranges of Refs. \cite{5,7} and their respective 
binding energies together with the  five-state result. 
We also
observe a correlation between  S-wave singlet Ps-H
binding and resonance energies, which is shown in Fig. 4. 
The essentially
exact resonance and binding energies  \cite{ho2,ho1} lie on
the  line in Fig. 4 obtained by our calculation and those of
Refs.
\cite{5,7}.  The reproduction of the correct value of the singlet 
Ps-H effective 
range and 
resonance energy in addition to the scattering length 
 assures proper variation of the phase shift in our
 model. The previous calculations \cite{5,5a,6,7} have  possibly
not converged well 
as 
they do not produce the correct energies of Ps-H bound state and
resonance.  
The  five-state 
model  reproduces the positions of the Ps-H bound state and resonance
fairly
accurately, and so it is expected  that the present singlet 
scattering length, 3.72$a_0$, effective range,
1.67$a_0$,  phase shifts and low-energy cross
sections are more close to the converged results than those of previous
calculations. The correlations of Fig. 2 and 3
suggest that the correct singlet
scattering length and effective range, corresponding to the accurate 
Ps-H binding energy of 1.067 eV \cite{ho1}, 
should be  3.5$a_0$ and 1.65$a_0$, respectively, in close agreement with
our model calculation. In the 
triplet case there is  no bound state and no
interesting correlation is observed. 

\vskip -2.3cm
\postscript{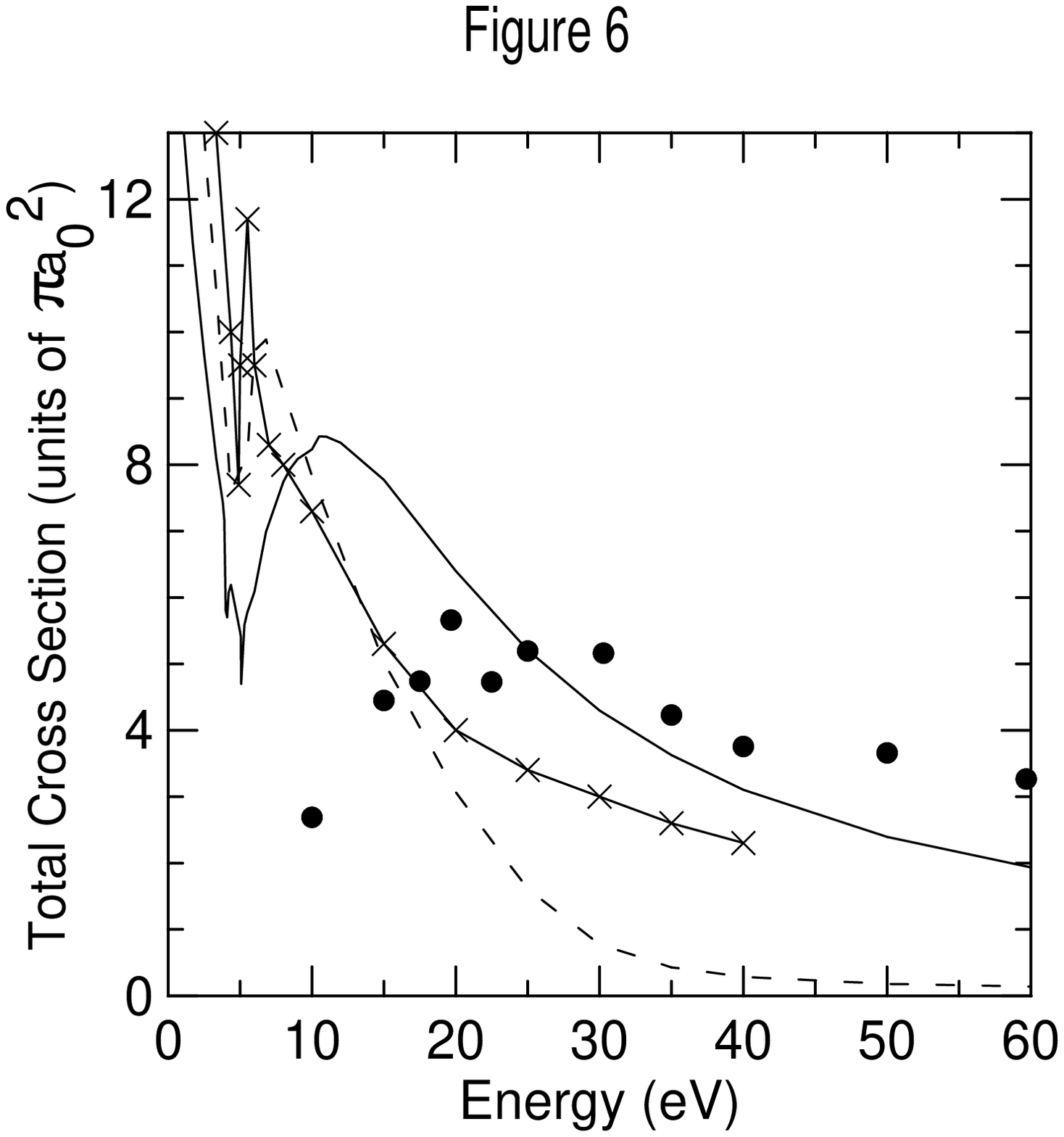}{1.0}    
\vskip -2.2cm
\noindent {\small {\bf Fig. 6} \ \
 Total  cross section 
for  Ps-H scattering at different 
Ps energies:  present  total (full line), 
three-Ps-state  
CCA  (dotted line, Ref. \cite{6}), target-elastic total of 22-pseudo-state
model including Ps-ionization and excitations
(full line, Ref. \cite{7}),
Ps-H$_2$ experiment
reduced by a factor of 2 ($\bullet$, Refs. \cite{2}).
}

\vskip .2cm {Table I: Singlet  scattering
length and effective range  in units
of $a_0$ and 
S-wave singlet binding energies ($E_B$) in eV for different
number of coupled states in two different models: Present and R-matrix
model of \cite{7}. The numbers with an asterisk denote prediction from
correlation of Figs. 2 and 3 and that with a dagger denote accurate
prediction in Ref. \cite{ho1}.  
} \vskip .2cm \begin{centering} \begin{tabular}
{|c|c|c|c|c|c|c|c|} \hline
&Ref. & 1-st   & 3-st  & 5-st & 9-st &22-st &
 \\
 \hline
$a^+$&Our & 4.05& 3.85 & 3.72& &   &  3.50$^*$ \\
$a^+$ &\cite{7}& 7.25& 6.70 & & 5.51& 5.20  &   \\
$r_0^+$&Our & 1.82& 1.72 & 1.67& &   &  1.65$^*$ \\
$r_0^+$ &\cite{7}& 3.07& 2.98 & &2.63 & 2.52  &   \\
$E_B$ &Our &0.87& 0.98& 1.05&  &  & 1.067$\dag$\\
$E_B$ & \cite{7}&0.263& 0.326& & 0.543 &0.634  & \\
\hline
\end{tabular}

\end{centering}
\vskip .2cm

To illustrate the trend of convergence of our calculation,
we show in
Table I the results for singlet scattering length, effective range, and
binding energies for one-,
three-, and five-state schemes with  model exchange potential  and compare
with the conventional  R-matrix calculation of
Ref. \cite{7} for different number of coupled states.  In this table we
also show the
predictions for the scattering length and effective range obtained from
correlations in Figs. 2 and 3 consistent with the correct Ps-H binding
\cite{ho1}. The triplet scattering lengths
 for the one-, three-, and
five-state models, which do not provide any
correlation,                                                 
 are 1.83$a_0$, 1.69$a_0$,  and 1.68$a_0$, respectively.   

The  model calculation leads to reasonable  
  convergence 
for cross section and phase shifts at low energies as the number of states
is increased. This is illustrated for low-energy  cross
sections in Table II for different basis sets.  Finally, we present the
low-energy phase shifts  and quenching cross sections of the
 five-state model   in Table III. As in Ref. \cite{7}, the quenching
cross section has a minimum between 0 and 1 eV and a maximum between 1 and 2
eV. However,  the  low-energy quenching as well as elastic cross
sections are somewhat smaller than those of Ref. \cite{7} and are expected to
be more converged.

\vskip .2cm {Table II: Low-energy  elastic cross
sections in units of $\pi a_0^ 2$ using different basis sets 
  for different $k$ in au,  incident positron energy $E=6.8 k^2$
eV.}
\vskip
.2cm \begin{centering} \begin{tabular} {|c|c|c|c|c|c|c|}
\hline
$k$  & 1-state   & 2-state  & 3-state & 4-state &
 5-state & 3-H-state\\
 \hline
0.0 & 26.39&24.78 &23.35& 23.35  &22.36  &  24.10 \\
0.1 &24.81 &23.44 &22.04 & 22.04&21.18&22.89 \\
0.2 & 21.24& 20.35&19.02 &19.00 &18.43& 19.89\\
0.3 &17.60 &17.13 &15.83 &15.83 &15.51& 16.79\\
0.4 &14.82 & 14.62&13.34 &13.34 &13.18& 14.42\\
0.5 & 12.76& 12.72&11.44 &11.43 &11.35& 12.58\\
0.6 & 11.01& 11.05& 9.72&9.72 &9.68& 10.89\\
0.7 & 9.42& 9.55  &8.14 &8.14 &8.10& 9.27\\
0.8 &8.02 & 7.69  & 6.20&6.19 &6.19& 7.82\\
\hline
\end{tabular}

\end{centering}

\vskip .2cm

In Fig. 5 we plot the   Ps(1s)H(1s), Ps(2s)H(1s), Ps(2p)H(1s),
Ps(1s)H(2s), and Ps(1s)H(2p) cross sections for  the 
five-state  calculation,  and the Born cross
sections for $n \ge 3$ Ps-excitations and Ps-ionization.
These cross sections are also  exhibited in Table IV.
The   total cross section 
is plotted in Fig. 6 where we 
 compare our results with
those of the  22-pseudo-state 
$R$-matrix and three-state CCA calculations of Refs. \cite{6,7}.
In the absence of experimental results
on Ps-H scattering, we compare the  total cross section with the total
Ps-H$_2$ cross section data ($\bullet$) reduced by a factor of two \cite{3}.
This should provide a fair comparison except at very low energies.   
The experimental trend, which 
 clearly demonstrates a broad maximum in the total 
cross section for all the Ps-impact scattering problems 
around  20  eV \cite{2}
and possibly a minimum near the Ps(2s) excitation
threshold \cite{ba2}, is correctly reproduced in our calculation.
The Ps-ionization cross section is
largely responsible for producing 
 this trend in Ps-H scattering and also
in Ps-He and Ps-H$_2$ scattering \cite{ba2,ba3}. 
The 22-pseudo-state calculations of Refs. 
\cite{7,8} do not have this trend even after including Ps-ionization and
H-excitation and ionization
cross sections. The Ps-H cross sections of Ref. \cite{6} shown 
in Fig. 5 do not include higher excitations and ionizations of Ps and H;
but the trend of their result suggests  that it
 may agree with the  experimental trend 
of a maximum if these cross sections are included.
However, 
  at low energies our cross sections are much smaller 
  than those of Refs. \cite{6,7}. 
\onecolumn  
  
{Table III: Low-energy phase shifts in radians and 
 ortho-Ps(1s) to para-Ps(1s) conversion cross
sections in units of $\pi a_0^  2$ for  the
 five-state
model  for different $k$ in au. The entries for $k=0$ correspond
to the scattering lengths in units of $a_0$, incident positron energy
$E=6.8 k^2$
eV.}
\vskip
.2cm

\begin{centering} \begin{tabular} {|c|c|c|c|c|c|c|c|}
\hline
$k$  & $\delta_0^{+}$   & $\delta_0^{-}$  & $\delta_1^{+}$ & 
$\delta_1^{-}$&  $\delta_2^{+}$ & $ \delta_2^{-}$  & $\sigma
_{\makebox{quen}}$\\
 (au) & (rad) &(rad)    &(rad) &(rad)&(rad)&(rad)&  
 ($\pi a_0 ^  2$) \\
 \hline
0.0 &3.72  &1.68  &  &  & & &1.02\\
0.1 & 2.78&$-$1.67($-$1) &4.77($-$3) & $-$2.33($-$3)&1.8($-$5) &$-$1.4($-$5)& 0.99\\
0.2 & 2.44&$-$3.27($-$1) &3.70($-$2) & $-$1.67($-$2)& 5.3($-$4)&$-$4.0($-$4) & 0.91\\
0.3 &2.14 &$-$4.74($-$1) &1.16($-$1) & $-$4.76($-$2)&3.5($-$3) &$-$2.6($-$3)& 0.93\\
0.4 & 1.89&$-$6.02($-$1) &2.39($-$1) & $-$9.18($-$2)&1.2($-$2)&$-$8.6($-$3) & 1.07\\
0.5 & 1.68&$-$7.06($-$1)& 3.72($-$1) &$-$1.42($-$1) &2.9($-$2) & $-$2.0($-$2) & 1.21\\
0.6 &1.52 &$-$7.84($-$1) &4.78($-$1) & $-$1.90($-$1)& 5.5($-$2) & $-$3.6($-$2)  &1.21\\
0.7 & 1.43 &$$7.373($-$1) &5.41($-$1) & $-$2.28($-$1)& 8.8($-$2)&$-$5.5($-$2)  &1.10\\
0.8 &4.12&$$7.196($-$1) & 5.69($-$1)&$-$2.47($-$1) &1.2($-$1) & $-$7.3($-$2) & 1.07\\
\hline
\end{tabular}

\end{centering}

\vskip .3 cm

  In this study we find that the H-excitation cross sections are 
much smaller than the Ps-excitation cross sections and the
H-excitation channels 
have  less effect on the convergence of the solution at low energies
compared to the Ps-excitation channels. This is consistent with the fact
that the polarizability of the H atom is one-eighth of that of the Ps
atom. 
In view of this, the difference between our  low-energy results 
 and those of Refs. \cite{6,7} seems to be due to  their unconverged
nature and not due to the neglect of H states in their model. 
This 
is  explicit  in their estimation for Ps-H binding energies.

\vskip .2cm {Table IV: Ps-H partial cross sections in
units of $\pi a_0^2$ at different positronium energies} \vskip
.2cm 
\begin{centering} \begin{tabular} {|c|c|c|c|c|c|c|c|} 
\hline
E & Ps(1s)- & Ps(2s)-  & Ps(2p)- & Ps(1s)- & Ps(1s)- 
 & Ps($n\ge 3$)- & Ps-ion- \\
(eV) & H(1s) & H(1s)  & H(1s) & H(2s) & H(2p) &
 H(1s) & H(1s)\\
  \hline
 5.08&4.70  &            &      & & & & \\ 
 5.5 & 4.88 & 1.01($-$1) & 7.92($-1$)& & & &  \\
 6 & 4.53&1.55($-$1) &1.40 & & & &  \\
 6.8& 4.10&1.52($-$1) &1.83 & & & 9.06($-$1)&   \\
8& 3.59&1.34($-$1) &1.93 & &  & 1.08 & 1.02  \\
10 & 2.91&1.19($-$1) &1.74 & & & 9.16($-$1)  & 2.55\\
11&2.64 &1.17($-$1) &  1.45   &1.38($-$1)  & 2.33($-$1)& 8.26($-1$) & 3.02   \\ 
12 & 2.34& 1.12($-$1)&1.28 & 2.06($-$1)& 3.04($-$1)&7.45($-$1)  &3.35 \\
15& 1.78&9.64($-$2) &1.07 & 1.99($-$1)& 3.10($-$1)   &5.60($-$1)  &3.76 \\
20& 1.19&6.91($-$2) &7.93($-$1) & 1.20($-$1)&2.11($-$1)&  3.79($-$1)  &3.64 \\
25& 8.17($-$1)&4.54($-$2) &5.98($-$1) & 7.48($-$2)& 1.37($-$1)&  2.78($-$1)   &3.26 \\
30& 5.73($-$1)&3.00($-$2) &4.68($-$1) & 4.88($-$2)& 9.09($-$2)& 2.16($-$1)   &2.87 \\
40& 3.08($-$1)&1.47($-$2) &3.16($-$1) & 2.42($-$2)&4.52($-$2) &  1.46($-$1)   &2.25 \\
60& 1.17($-$1)&5.11($-$3) &1.85($-$1) & 8.71($-$3)& 1.57($-$2)&  8.50($-$2)  &1.52 \\
80& 5.65($-$2)&2.37($-$3) &1.28($-$1) & 4.08($-$3)& 7.20($-$3)& 5.90($-$2)    & 1.13\\
100& 3.15($-$2)&1.29($-$3) &9.72($-$2) & 2.22($-$3)& 3.87($-$3)& 4.50($-$2)    &0.90 \\
\hline
\end{tabular}

\end{centering}

\section{Summary}

We have performed a five-state calculation of Ps-H scattering using a
recently proposed non-local model exchange potential.  The model considers
excitation of both Ps and H atoms and yields cross sections for
transitions to following final states starting from the initial state
Ps(1s)H(1s): Ps(1s)H(1s), Ps(2s)H(1s), Ps(2p)H(1s), Ps(1s)H(2s), and
Ps(1s)H(2p).  Higher excitations and ionization of the Ps atom are treated
by the Born approximation including  exchange.  The
cross sections are in qualitative agreement with experimental trend.  Our
 five-state model  yields singlet and triplet scattering
lengths of 3.72$a_0$  and 1.68$a_0$, and the singlet effective range
of $1.67a_0$.  The  calculation  reproduces the singlet S-wave Ps-H
resonance at 4.01 eV \cite{ho2} and predicts a Ps-H binding energy of
$1.05$
eV
compared to the accurate binding energy of   1.067 eV \cite{ho1}. This
assures us as to
the realistic nature of our model.  We observe  correlations
between the S-wave singlet Ps-H binding energy and 
singlet  scattering length, effective range, and  resonance
energy obtained  in
different calculations. These correlations of other observables with
binding energy 
demostrate the degree of
convergence of 
various model calculations as can be seen  in Figs. 2, 3,
and 4. 
Considering precise Ps-H binding energy of 1.067
eV,   correlations in Figs. 2 and 3 suggest a singlet scattering
length   of 3.5$a_0$ and an effective range of 1.65$a_0$.
The inclusion of higher-order  states in our five-state model are not
expected  
to influence the low-energy results significantly as has been demonstrated in
Table II, but  their effect could
be considerable  
at medium to high energies. A further detailed 
 calculation including these states
will help to understand the dynamics more precisely.

The work is supported in part by the Conselho Nacional de Desenvolvimento -
Cient\'\i fico e Tecnol\'ogico,  Funda\c c\~ao de Amparo
\`a Pesquisa do Estado de S\~ao Paulo,  and Finan\-ciadora de Estu\-dos e
Projetos of Brazil.


\begin{references}
\bibitem{xx} This is in fact $-2\pi^ 2$ times the model exchange potential or
the exchange Born term.
\end{references}
\end{document}